\documentclass[acmsmall,screen,nonacm]{acmart}

\usepackage{soul}
\usepackage{xcolor}
\usepackage{graphicx}
\sethlcolor{pink}

\AtBeginDocument{%
  }

\setcopyright{acmlicensed}
\copyrightyear{2026}
\acmYear{2026}
\acmDOI{XXXXXXX.XXXXXXX}
\acmConference[Conference acronym 'XX]{Make sure to enter the correct
  conference title from your rights confirmation email}{June 03--05,
  2018}{Woodstock, NY}
\acmISBN{978-1-4503-XXXX-X/2018/06}

\begin{document}


\title[From Chaos to Clarity]{From Chaos to Clarity: A Framework for Program-Level AI Learning Outcomes}


\author{Grace Barkhuff}
\affiliation{%
  \institution{Georgia Institute of Technology}
  \city{Atlanta}
  \state{Georgia}
  \country{USA}}
\email{grace.barkhuff@gatech.edu}

\author{Ian Pruitt}
\affiliation{%
  \institution{Georgia State University}
  \city{Atlanta}
  \state{Georgia}
  \country{USA}}
\email{ipruitt2@gsu.edu}

\author{William Gregory Johnson}
\affiliation{%
  \institution{Georgia State University}
  \city{Atlanta}
  \state{Georgia}
  \country{USA}}
\email{wjohnson6@gsu.edu}

\author{Rodrigo Borela}
\affiliation{%
  \institution{Georgia Institute of Technology}
 \city{Atlanta}
  \state{Georgia}
  \country{USA}}
\email{rborelav@gatech.edu}

\author{Ben Rydal Shapiro}
\affiliation{%
  \institution{Georgia State University}
  \city{Atlanta}
  \state{Georgia}
  \country{USA}}
\email{bshapiro@gsu.edu}

\author{Anu G. Bourgeois}
\affiliation{%
  \institution{Georgia State University}
  \city{Atlanta}
  \state{Georgia}
  \country{USA}}
\email{abourgeois@gsu.edu}

\renewcommand{\shortauthors}{Barkhuff et al.}

\begin{abstract}
Industry is leaning into generative artificial intelligence (GenAI), and higher education is under pressure to prepare graduates for a GenAI-augmented workforce. Yet, there is still no clear structure for defining AI readiness across disciplines, programs, courses, and assignments. Current approaches often rely on broad institutional policies or individual course-level decisions, which can also create mixed messages for students, fragmented expectations across programs, and limited visibility for university leaders. In this paper, we argue that higher education needs a more coherent way to connect institutional priorities to curriculum-level action. We propose Program-Level AI Learning Outcomes (PLAI-LOs) as a framework for defining what students graduating from a program should know and be able to do with, without, and about GenAI in a given discipline. The PLAI-LOs framework complements existing program-level learning outcomes and supports alignment across institutional priorities, program-level AI learning outcomes, course-level learning outcomes, and assignment-level objectives. We illustrate the framework with examples from computing and music and show how PLAI-LOs can be implemented through artifact-level GenAI policies, helping programs decide where GenAI should be taught and used, and when students should be expected to work without GenAI. We offer PLAI-LOs as a concrete, measurable, and adaptable path for moving higher education from scattered GenAI rules toward a strategy with clear, learning-centered alignment.

\end{abstract}




\maketitle

\section{Introduction}

One thing is clear: industry is leaning into AI, and employers increasingly expect workers to be proficient with generative AI (GenAI) tools \cite{Gatta_2026_AI_skills, Sigelman_2025_GenAI_learningcurves}.
Higher education is trying to respond by hosting summits, forming task forces, developing AI policies, and asking programs how they are preparing students for a future shaped by GenAI \cite{JIN_2025GenAI_Theory, niklas_2025_Higher_AIpolicies}.
Yet, amid all of this activity, there is still uncertainty about what steps to take.
What does it mean for a graduate to be AI-ready? Who decides? Is it the same for all students regardless of discipline?
While higher education is moving quickly to respond to GenAI, it does not yet have a clear structure for turning that urgency into curriculum-level action \cite{ma2025preparing}.

That lack of structure is creating pressure across the system. Administrators want to report that their programs are producing graduates ready for the workforce \cite{voss_2026_AI_workforce}. Faculty are being pushed to adjust quickly while still protecting the learning they are responsible for supporting \cite{ma2025preparing}. At the same time, faculty members themselves are trying to learn how to use AI \cite{gorbunova_2026_barriers_AI_adoption, arroyo2025generative}.
Students are watching job roles change or disappear in real time, especially at the entry level, and many are asking whether the degree they are pursuing will prepare them for the workforce they expected to enter \cite{NACE2025, WorldEconomic2025, Bay_2026_AI_Skills}. 
This anxiety is showing up directly in classrooms  and leaving both groups with unresolved questions \cite{farooqi2026jobanxietypostsecondarycomputer, kasneci2023chatgpt, cotton2024chatting}:
When should GenAI be introduced, when should it be limited, and what should students be able to do without it?
If GenAI can do part of the work, what needs to be learned?
If students use GenAI to complete required work, is this better preparation for the workforce, or does it undermine future success by not learning foundational concepts and skills? 

This course-by-course and individual faculty approach is problematic -- decisions about GenAI are often being made at disconnected levels.
While it provides local flexibility and instructor autonomy, it can also produce mixed messages for students and fragmented expectations across a program \cite{cotton2024chatting}. 
Institutional GenAI policies and general syllabus statements are important, but they cannot fully specify what GenAI use should look like within every discipline, program, course, or learning artifact.
This is especially true when the appropriate use of GenAI may vary across homework assignments, labs, exams, quizzes, group projects, and other forms of assessment.

We argue that higher education needs a clearer structure between broad institutional guidance and individual course policies. 
The central question should not only be whether GenAI is allowed, but what students graduating from a program should know and be able to do with, without, and about GenAI.
This question is beginning to be addressed through calls for program-level AI planning and institutional efforts to map AI competencies into curricula \cite{Landy_2026_Prog-level_AI, miamioh_2026_AI_majors}. 
We build upon this emerging direction by proposing a concrete framework for connecting program-level AI expectations to course-level outcomes, assignment-level objectives, and university-level guidance.

In this paper, we propose Program-Level AI Learning Outcomes (PLAI-LOs) as one way to create that structure. 
They are meant to complement existing program-level learning outcomes (PLOs), not replace them.
PLOs describe the broader knowledge and skills students should acquire that are particular to a given discipline.
PLAI-LOs are program-level statements that clarify what students graduating from a program should know and be able to do \textit{with, without, and about GenAI} for a given discipline. This is not simply AI-literacy, but specific measurable outcomes for graduating students.

Our contribution of PLAI-LOs is a framework that moves these conversations toward actionable steps. 
The framework connects four levels (and stakeholders) that are often treated separately: institutional priorities (university administrators), program-level AI learning outcomes (program administrators), course-level learning outcomes (faculty), and assignment-level objectives (students). At each level, the outcomes should be SMART which we define as specific, measurable, actionable, relevant, and time-constrained, so that AI-readiness can be defined, implemented, assessed, and revisited as the technology changes and specific fields adapt.

\section{The Framework}

\begin{figure*}
    \centering
    \includegraphics[width=\linewidth]{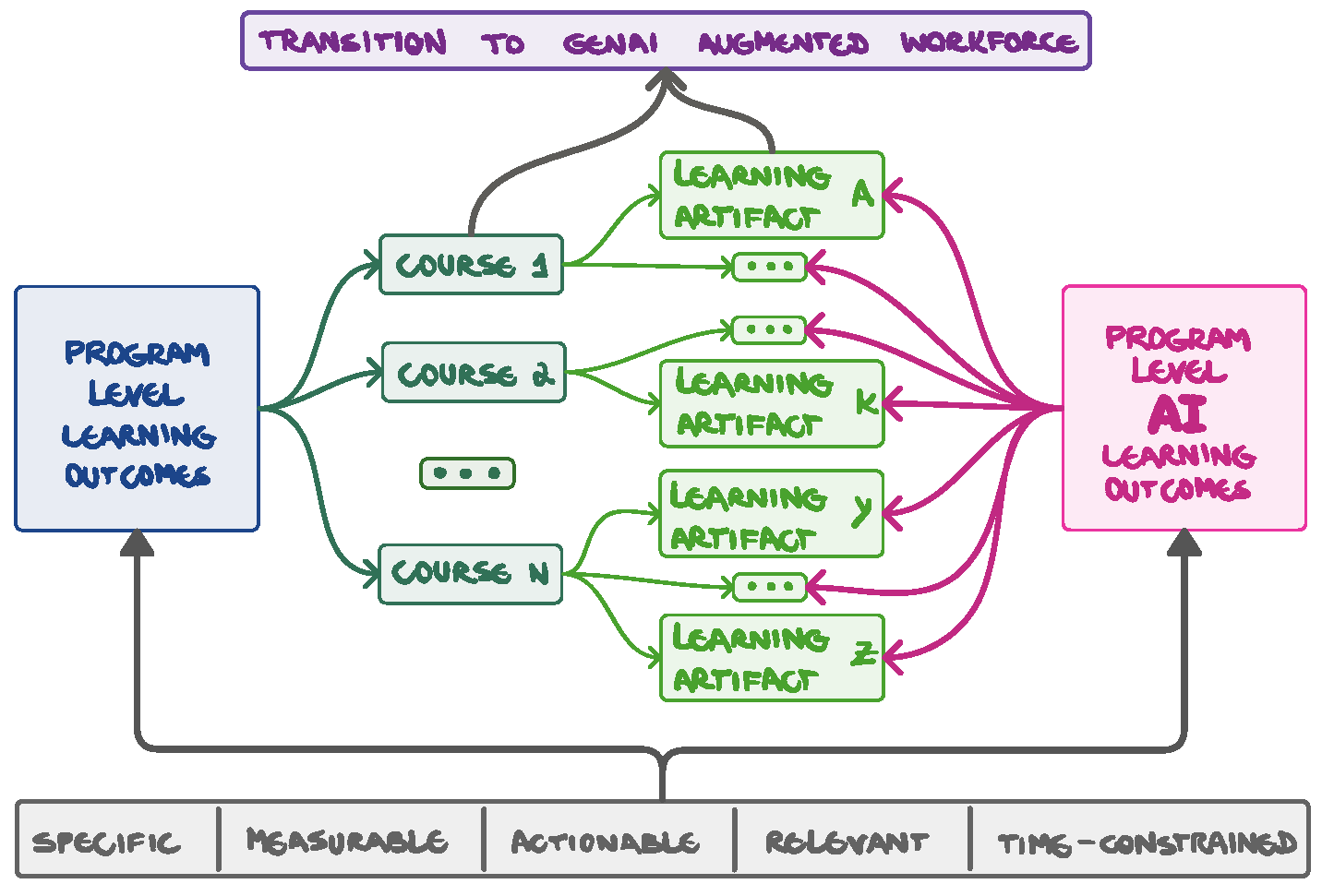}
    \caption{The framework develops program-level GenAI learning outcomes, maps them to learning artifacts, and implements them through artifact-level GenAI use policies.}
    \label{fig:framework}
\end{figure*}

The framework includes two main ideas: (1) defining Program-Level AI Learning Outcomes (PLAI-LOs) through a Backward Design process \cite{wiggins_2005_understanding}, and (2) implementing the PLAI-LOs via individual learning artifact GenAI policies. Together, the framework strengthens alignment between program administrators, instructors, and students on how and why GenAI should be used across any individual program's coursework. Furthermore, by establishing a mechanism to monitor PLAI-LO implementation, administrators can assess the progress of curriculum redesign initiatives. The framework is visualized in Figure \ref{fig:framework}, which shows how traditional program learning outcomes (PLOs) are mapped to courses and PLAI-LOs are mapped to individual course artifacts. Both are built upon ``SMART'' learning outcomes, and lead to students transitioning successfully into a GenAI-augmented workforce.

\subsection{PLAI-LOs: Design Process and Assessment}

Given the fast evolving nature of GenAI, the design, implementation, and assessment of PLAI-LOs should be recurring and iterative, with a time frame aligned with institutional calendars to avoid cycles of confusion among faculty and students. The iterative cycle of the framework is described in the following.

\textit{Step 1: } First, institutions should assess or re-assess the state-of-the-art of GenAI technologies and the state-of-practice in their programs to identify emerging roles and skills expected from graduates to be successful in their careers.

\textit{Step 2: } Utilizing a Backward Design Process \cite{wiggins_2005_understanding}, program leaders will meet to translate the new developments in GenAI technology and practice into PLAI-LOs, utilizing the revised Bloom's taxonomy \cite{krathwohl_2002_bloom}. These PLAI-LOs represent student learning outcomes across the entire program and should be relatively broad.

\textit{Step 3: } Here, instructors will take the PLAI-LOs and apply them to their own courses. Each instructor can design or update formative and summative assessments (learning artifacts) to integrate (or not) GenAI components that match the targeted PLAI-LOs. These learning artifacts then receive a tag corresponding to those PLAI-LOs, which can be used to report which PLAI-LOs are being met in which artifact.

\textit{Step 4: } Instructors develop artifact-level GenAI policies, which delineate what is considered acceptable use in the context of each individual learning artifact to ensure pedagogical rigor and alignment with the targeted PLAI-LOs.

\textit{Step 5: } Finally, student success on the corresponding learning artifacts with a PLAI-LO tag is reported back to program leaders and evaluators to enable monitoring of the implementation across the curriculum. The cycle repeats from step 1 in the following academic term.

This framework offers various efficiencies from both curriculum design and program evaluation perspectives. Since the PLAI-LOs parallel the existing learning outcomes, they reduce friction associated with changing the existing program-level learning outcomes and work to augment rather than replace foundational learning. By tying PLAI-LOs directly to learning artifacts, instructors can make gradual interventions rather than undertake full course redesigns, which may become outdated quickly given the rapid pace of GenAI development. Artifact-level GenAI policies provide clarity to students on what and why they are learning, both in terms of GenAI skills and as it pertains to their majors. Finally, the use of tags and reporting enables easy monitoring and quantification of PLAI-LO implementation progress for program evaluators.

\subsection{PLAI-LOs: Implementation Examples}

Based on the current state of GenAI technologies and practice in different professions, we provide an example of six PLAI-LOs written for an undergraduate computer science (CS) program and for an undergraduate music program (See Table \ref{tab:GLOs}). These are just examples, inspired by our background in computer science education, one author's music background, the music program learning outcomes from Adam's State University \cite{noauthor_music_nodate}, and a review article on the use of GenAI in music education \cite{mazlan_artificial_2026}.

\begin{table*}[]
    \centering
    \begin{tabular}{p{.45\linewidth}|p{.45\linewidth}}
        \textbf{Sample PLAI-LOs for Computing}& \textbf{Sample PLAI-LOs for Music}\\ \hline
        \textbf{(1) Demonstrate core CS proficiency without GenAI:} A CS graduate can independently design, implement, test, and debug multi-file software using appropriate data structures, algorithms, and modular design—without relying on GenAI. & \textbf{(1) Demonstrate high proficiency in performance, composition, and conducting without GenAI:} Music graduates are able to express themselves artistically at a high level without any GenAI involvement.\\ \hline
        \textbf{(2) Communicate computing work clearly and technically without GenAI:} A CS graduate can explain system design, algorithmic choices, implementation tradeoffs, and results through technical writing and oral presentation, in their own words, without GenAI dependence.& \textbf{(2) Use GenAI for advanced writing and brainstorming:} Music graduates will be able to express themselves in written form without the use of GenAI, but are also knowledgeable about how to use GenAI for advanced writing.\\ \hline
        \textbf{(3) Use GenAI for early-stage problem framing, then engineer the solution:} A CS graduate can use GenAI to brainstorm approaches, compare alternatives, and clarify requirements, then independently translate those ideas into working code and system components. & \textbf{(3) Understand the ethical concerns around GenAI use by the artistic community:} Music graduates will be prepared to discuss and make choices about using GenAI in their own music workflow while understanding the ethical concerns held by the artistic community as a whole.\\ \hline
        \textbf{(4) Build and debug non-trivial software in GenAI-augmented workflows:} A CS graduate can effectively use AI-assisted development tools to produce larger, more complex software systems, while maintaining code quality, test coverage, maintainability, and performance. & \textbf{(4) Co-compose with GenAI:} Music graduates can use GenAI as a composition collaborator for advanced, multi-instrument pieces. \\ \hline
        \textbf{(5) Validate AI-assisted code and outputs with CS rigor:} A CS graduate can critically evaluate GenAI-generated code and explanations for correctness, security, efficiency, and robustness via testing, static/dynamic analysis, and systematic debugging. & \textbf{(5) Use GenAI as a practice assessor:} Music graduates are able to use AI tools to evaluate their practice session, explain the accuracy of the evaluation, and prepare a practice plan to address any realistic issues brought up by the AI.\\ \hline
        \textbf{(6) Make disciplined, ethical, and professional GenAI decisions in computing contexts:} A CS graduate can identify and address risks in GenAI-supported outputs, e.g., bias, privacy, IP/licensing, security, misuse, and over-reliance; and justify when, where, and how GenAI should or should not be used. & \textbf{(6) Evaluate AI-generated musical material:} Music graduates can critically assess AI-generated musical material for quality, stylistic fit, and alignment with artistic intent. \\ \hline
    \end{tabular}
    \caption{Example PLAI-LOs for a CS program (L) and music program (R).}
    \label{tab:GLOs}
\end{table*}

\subsubsection{CS PLAI-LO Learning Artifact Example}

To make the framework concrete, we consider the implementation of the fourth Computing PLAI-LO (Build and debug software with GenAI) in a reinforcement learning (RL) course artifact. An upper-level RL course may have as one of its summative assessments, a project in which students solve the Lunar Lander environment with a continuous action space \cite{towers_2025_gymnasium}. To ensure students develop a deep understanding of the mechanics of policy-gradient methods, the GenAI policy may require that students implement one algorithm from scratch (e.g. PPO \cite{schulman_2017_ppo}) without any GenAI assistance. As part of the assignment, however, students may also be expected to compare the performance of their chosen and implemented algorithm with two others (e.g. TD3 \cite{fujimoto_td3_2018} and SAC \cite{haarnoja_2028_sac}). In that instance, the GenAI policy for that artifact may allow students to build these models with GenAI assistance via coding agents or collaboration with a chatbot to accelerate the process, since the learning outcome associated with this component of the artifact relates to evaluation of different techniques.

\subsubsection{Music PLAI-LO Learning Artifact Example}

Let us also consider a concrete example within the music program. Consider an undergraduate introductory music composition course (a course on writing music). To meet the first PLAI-LO (Demonstrate high proficiency in performance, composition, and conducting without GenAI), most of the course will be taught where GenAI use on artifacts and assignments is prohibited. This is made clear to the students, and they are also told that at the end of the course there will be one assignment where GenAI use is required.

That assignment will implement PLAI-LO four (Co-compose with GenAI). For the assignment, students could be asked to use GenAI as a collaborator to write music. The assignment might scaffold different steps in the collaboration process, from brainstorming the theme, feel, and instrumentation of the music to the actual writing of melody and harmony. Finally, students may submit a written reflection on the process, including their attitudes toward the collaboration and opinions on the final musical output (which begins to address PLAI-LO six: Evaluate AI-generated musical material). This assignment serves as an introduction to the GenAI music collaboration process, and later in the curriculum, such as in an advanced composition course, students may have a deeper collaboration with GenAI to continue developing PLAI-LO four.
\section{Discussion}

In this paper, we outline a process for achieving GenAI alignment across various university stakeholders, from university administrators to students, by using Program-Level AI Learning Outcomes (PLAI-LOs). PLAI-LOs, which parallel existing program learning outcomes, allow programs to define what students should know and be able to do with, without, and about GenAI. They help instructors determine if, when, and how students should use GenAI or abstain from its use, and articulate this clearly and consistently to students. They also allow programs to measure GenAI use and learning and report this concretely to those in university leadership. Additionally, the framework provides cohesion not just up and down the university hierarchy, but also across differing programs. When students take courses in different departments or university leadership assesses various programs, the PLAI-LOs framework provides consistency across those experiences.

Because the PLAI-LOs framework uses the existing learning outcomes model, the process is ``SMART'' (specific, measurable, actionable, relevant, and time-constrained). The framework prepares programs to outline specific tasks students should be able to do or things students should know with and without GenAI by the time they graduate,  measure and report progress, set reasonable standards for students, connect outcomes to the specific program, and update expectations as AI technology changes.

However, we invite and encourage commentary from all stakeholders: AI researchers and technologists, educators, university leaders, and students. Our goal is to outline a framework and process that is customizable to any program, regardless of discipline or institution type, and we need feedback in order to improve upon and solidify the process. For example, we wonder: \textit{is this framework clear for students? Will it have instructor buy-in? Is it sufficiently measurable for university leadership?}


We also make suggestions for a continued and future research agenda. While research is beginning to understand how GenAI tools for pedagogical use impact student learning \cite{vieriu2025impact, Pallant03042026}, 
our framework describes \textit{what skills and content knowledge} students need to know about and be able to do with GenAI to prepare them for the workforce and a future where GenAI is commonplace. This goes beyond basic AI literacy, toward integration of the tool into daily work. Research is needed to understand how this kind of GenAI-related disciplinary knowledge can be defined, taught, and assessed across programs.

Additionally, we hypothesize that clear and cohesive learning outcomes for students, presented to them on individual artifacts, will provide students with a sense of agency in their own learning. Research should examine whether artifact-level guidance increases students' engagement with their own learning, rather than outsourcing it to GenAI in a prohibited way \cite{walsh_everyone_2025}.

\section{A Call to Action}
GenAI use in higher education is at a critical juncture. The current per-course and per-instructor approach is confusing to students, difficult for instructors, and insufficient for university administrators. 
Higher education needs structures that help programs decide what students should learn, where GenAI belongs, when to intentionally refrain from use of GenAI, and how those decisions should be revisited as the technology changes.
The PLAI-LOs framework is implementable, measurable, and aligned with needs of stakeholders across higher education. 
If higher education wants to prepare students for a GenAI-augmented workforce, then programs need more than scattered rules; they need clear, measurable, and learning-centered structures that move GenAI use from confusion to purposeful preparation.










\begin{acks}
Research was sponsored in part by the Army Research Laboratory and was accomplished under Cooperative Agreement Number W911NF-23-2-0224 and in part by Google. The views and conclusions contained in this document are those of the authors and should not be interpreted as representing the official policies, either expressed or implied, of the Army Research Laboratory or the U.S. Government. The U.S. Government is authorized to reproduce and distribute reprints for Government purposes notwithstanding any copyright notation herein.
\end{acks}


\bibliographystyle{ACM-Reference-Format}
\bibliography{bibliography.bib}

@misc{Bay_2026_AI_Skills,
    author = {Jordan Bay} ,
    title = {As AI Skills Surge, Entry-Level Jobs Lag},
    url = {https://www.insidehighered.com/news/student-success/life-after-college/2026/04/30/ai-skills-surge-entry-level-jobs-lag?utm_source=openai},
    urldate = {2026-06-29},
    Journal = {Inside Higher Ed},
    year = {2026}
}

@article{cotton2024chatting,
  title={Chatting and cheating: Ensuring academic integrity in the era of ChatGPT},
  author={Cotton, Debby RE and Cotton, Peter A and Shipway, J Reuben},
  journal={Innovations in education and teaching international},
  volume={61},
  number={2},
  pages={228--239},
  year={2024},
  publisher={Taylor \& Francis}
}

@article{kasneci2023chatgpt,
  title={ChatGPT for good? On opportunities and challenges of large language models for education},
  author={Kasneci, Enkelejda and Se{\ss}ler, Kathrin and K{\"u}chemann, Stefan and Bannert, Maria and Dementieva, Daryna and Fischer, Frank and Gasser, Urs and Groh, Georg and G{\"u}nnemann, Stephan and H{\"u}llermeier, Eyke and others},
  journal={Learning and individual differences},
  volume={103},
  pages={102274},
  year={2023},
  publisher={Elsevier}
}

@article{ma2025preparing,
  title={Preparing students for an AI-driven world: Generative AI and curriculum reform in higher education},
  author={Ma, Ying and Su, Youxiang and Li, Mingda and Zhang, Yu and Chai, Wantong and Huang, Amin and Zhao, Xiaofei},
  journal={Frontiers of Digital Education},
  volume={2},
  number={4},
  pages={30},
  year={2025},
  publisher={Springer},
  date = {2025-09-15}
}

@article{niklas_2025_Higher_AIpolicies,
author = {Humble, Niklas},
title = {Higher Education AI Policies—A Document Analysis of University Guidelines},
journal = {European Journal of Education},
volume = {60},
number = {3},
pages = {e70214},
doi = {https://doi.org/10.1111/ejed.70214},
url = {https://onlinelibrary.wiley.com/doi/abs/10.1111/ejed.70214},
note = {e70214 5877432},
year = {2025},
date = {2025-08-21}
}

@article{JIN_2025GenAI_Theory,
    title = {Generative AI in higher education: A global perspective of institutional adoption policies and guidelines},
    journal = {Computers and Education: Artificial Intelligence},
    volume = {8},
    pages = {100348},
    year = {2025},
    issn = {2666-920X},
    doi = {10.1016/j.caeai.2024.100348},
    author = {Yueqiao Jin and Lixiang Yan and Vanessa Echeverria and Dragan Gašević and Roberto Martinez-Maldonado}
}

@misc{Landy_2026_Prog-level_AI,
    author = {Kathleen Landy},
    title = {The Program-Level AI Conversations We Should Be Having},
    url = {https://www.insidehighered.com/opinion/views/2024/02/28/next-step-higher-eds-approach-ai-opinion},
    journal = {Inside Higher Ed},
    date = {2024-02-28},
    urldate = {2026-06-29},
    year = {2026}
}

@techreport{miamioh_2026_AI_majors,
    author = {Board of Trustees},
    title = {Academic and Student Affairs Committee Report},
    institution = {Miami University at Oxford, OH},
    url = {https://miamioh.edu/about/leadership-administration/_files/documents/bot/2026/agenda-05-26-asa.pdf},
    date = {2026-05-14},
    urldate = {2026-06-30},
    year = {2026},
    pages = {204}
}

@techreport{voss_2026_AI_workforce,
    author = {Clara Voss PhD},
    title = { Hearing: Building an AI-Ready America: Higher Education in the Age of AI.},
    institution = {U.S. House Education and Workforce Subcommittee on Higher Education and Workforce Development},
    url = {https://www.academicjobs.com/us/higher-education-news/congress-hearing-on-ai-ready-workforce-in-higher-education-or-academicjobs-22408},
    date = {2026-06-03},
    urldate = {2026-06-30},
    year = {2026}
}

@ARTICLE{gorbunova_2026_barriers_AI_adoption,
	AUTHOR={Gorbunova, Nadezhda and 
			Vorsin, Eduard and 
			Spirina, Yelena and 
			Zhumagulova, Saule and 
			Popova, Nadezhda },   
	TITLE={Barriers to artificial intelligence adoption among instructors of IT-related disciplines in higher education: a survey-based study},
	JOURNAL={Frontiers in Education},
	VOLUME={Volume 11 - 2026},
	YEAR={2026},
	URL={https://www.frontiersin.org/journals/education/articles/10.3389/feduc.2026.1804254},
	DOI={10.3389/feduc.2026.1804254},
	ISSN={2504-284X}}

@article{arroyo2025generative,
  title={Generative AI and academic scientists in US universities: Perception, experience, and adoption intentions},
  author={Wenceslao Arroyo-Machado and 
          Jinghuan Ma and 
          Tipeng Chen and 
          Timothy P Johnson and 
          Shaika Islan and 
          Lesley Michalegko and 
          Eric Welch},
  journal={PloS one},
  volume={20},
  number={8},
  pages={e0330416},
  year={2025},
  publisher={Public Library of Science San Francisco, CA USA}
}

@misc{farooqi2026jobanxietypostsecondarycomputer,
      author = {Daniyaal Farooqi and 
                Gavin Pu and 
                Shreyasha Paudel and 
                Sharifa Sultana and 
                Syed Ishtiaque Ahmed},
      title = {Job Anxiety in Post-Secondary Computer Science Students Caused by Artificial Intelligence}, 
      year = {2026},
      archivePrefix={arXiv},
      primaryClass={cs.CY},
      url={https://arxiv.org/abs/2601.10468},
      date = {2026-01-15},
      urldate ={2026-06-29}
}

@misc{Sigelman_2025_GenAI_learningcurves,
    author = {Matt Sigelman and Joseph Fuller and Michael Fenlon and Erik Leiden and Gwynn Guilford},
    title = {The Expertise Upheaval, How Generative AI’s Impact on Learning Curves Will Reshape the Workplace},
    year = {2025},
    journal = {Harvard Business School and Burning Glass Institue},
    url = {https://www.hbs.edu/ris/Publication%20Files/The%20Expertise%20Upheaval%20-%20How%20GenAI%20Will%20Reshape%20the%20Workplace%20(25.07.09)_8f120802-d4d2-44de-97ed-94e81448a9de.pdf},
    urldate = {2026-06-30}
}

@misc{Gatta_2026_AI_skills,
    author = {Mary Gatta PhD},
    title = {Demand for AI Skills in Entry-level Jobs Nearly Triples Since Fall 2025 (NACE)}, 
    year = {2026},
    journal = {National Association of Colleges and Employers},
    url = {https://www.naceweb.org/job-market/trends-and-predictions/demand-for-ai-skills-in-entry-level-jobs-nearly-triples-since-fall-2025},
    urldate = {2026-06-29}
}

@misc{WorldEconomic2025,
title={The Future of Jobs Report 2025 World Economic Forum},
author={World Economic Forum},
url= {https://www.weforum.org/publications/the-future-of-jobs-report-2025/},
urldate = {2026-02-14},
}

@misc{NACE2025,
title= {Job Outlook 2025 (Revised january 2025; Spring Update 2025)},
author= {National Association of Colleges and Employers (NACE)},
url =  {https://www.naceweb.org/docs/default-source/default-document-library/2025/publication/research-report/2025-nace-job-outlook-jan-2025.pdf},
urldate = {2026-02-14}
}

@article{mazlan_artificial_2026,
	title = {Artificial intelligence applications and pedagogical challenges in music education},
	volume = {5},
	issn = {2731-5525},
	url = {https://doi.org/10.1007/s44217-026-01127-3},
	number = {1},
	journal = {Discover Education},
	author = {Mazlan, Chamil Arkhasa Nikko and Hanafi, Hafizul Fahri and Sarifin, Muhammad Ridhwan and Md Noor, Ahmad Rithaudin and Sadykova, Saule  Altynbayevna and Hidayatullah, Riyan and Jamnongsarn, Surasak},
	month = jan,
	year = {2026},
	pages = {140},
}

@misc{noauthor_music_nodate,
	title = {Music {Learning} {Outcomes} - {Music}},
	url = {https://www.adams.edu/academics/undergraduate/music/learning-outcomes/},
	language = {en-US},
	urldate = {2026-06-30},
	journal = {Adams State University},
}

@book{wiggins_2005_understanding,
  title={Understanding by Design},
  author={Wiggins, Grant and McTighe, Jay},
  edition={Expanded 2nd},
  year={2005},
  publisher={Association for Supervision and Curriculum Development},
  address={Alexandria, VA},
  isbn={978-1-4166-0035-0}
}

@article{krathwohl_2002_bloom,
author = {David R. Krathwohl and},
title = {A Revision of Bloom's Taxonomy: An Overview},
journal = {Theory Into Practice},
volume = {41},
number = {4},
pages = {212--218},
year = {2002},
publisher = {Routledge},
URL = {https://doi.org/10.1207/s15430421tip4104_2},
}

@misc{towers_2025_gymnasium,
title={Gymnasium: A Standard Interface for Reinforcement Learning Environments}, 
author={Mark Towers and Ariel Kwiatkowski and Jordan Terry and John U. Balis and Gianluca De Cola and Tristan Deleu and Manuel Goulão and Andreas Kallinteris and Markus Krimmel and Arjun KG and Rodrigo Perez-Vicente and Andrea Pierré and Sander Schulhoff and Jun Jet Tai and Hannah Tan and Omar G. Younis},
year={2025},
archivePrefix={arXiv},
primaryClass={cs.LG},
url={https://arxiv.org/abs/2407.17032}, 
}

@article{schulman_2017_ppo,
  author       = {John Schulman and
                  Filip Wolski and
                  Prafulla Dhariwal and
                  Alec Radford and
                  Oleg Klimov},
  title        = {Proximal Policy Optimization Algorithms},
  journal      = {CoRR},
  volume       = {abs/1707.06347},
  year         = {2017},
  url          = {http://arxiv.org/abs/1707.06347},
  eprinttype   = {arXiv},
  bibsource    = {dblp computer science bibliography, https://dblp.org}
}

@article{fujimoto_td3_2018,
  author       = {Scott Fujimoto and
                  Herke van Hoof and
                  David Meger},
  title        = {Addressing Function Approximation Error in Actor-Critic Methods},
  journal      = {CoRR},
  volume       = {abs/1802.09477},
  year         = {2018},
  url          = {http://arxiv.org/abs/1802.09477},
  eprinttype   = {arXiv},
  bibsource    = {dblp computer science bibliography, https://dblp.org}
}

@article{haarnoja_2028_sac,
  author       = {Tuomas Haarnoja and
                  Aurick Zhou and
                  Kristian Hartikainen and
                  George Tucker and
                  Sehoon Ha and
                  Jie Tan and
                  Vikash Kumar and
                  Henry Zhu and
                  Abhishek Gupta and
                  Pieter Abbeel and
                  Sergey Levine},
  title        = {Soft Actor-Critic Algorithms and Applications},
  journal      = {CoRR},
  volume       = {abs/1812.05905},
  year         = {2018},
  url          = {http://arxiv.org/abs/1812.05905},
  eprinttype   = {arXiv},
  bibsource    = {dblp computer science bibliography, https://dblp.org}
}

@article{vieriu2025impact,
  title={The impact of artificial intelligence (AI) on students’ academic development},
  author={Vieriu, Aniella Mihaela and Petrea, Gabriel},
  journal={Education Sciences},
  volume={15},
  number={3},
  pages={343},
  year={2025},
  publisher={MDPI}
}

@article{Pallant03042026,
author = {Jessica L. Pallant and 
		Janneke Blijlevens and 
		Alexander Campbell and 
		Ryan Jopp},
title = {Mastering knowledge: the impact of generative AI on student learning outcomes},
journal = {Studies in Higher Education},
volume = {51},
number = {4},
pages = {714--735},
year = {2026},
publisher = {Routledge},
doi = {10.1080/03075079.2025.2487570},
URL = {https://doi.org/10.1080/03075079.2025.2487570}
}

@misc{walsh_everyone_2025,
	title = {Everyone {Is} {Cheating} {Their} {Way} {Through} {College}},
	url = {https://nymag.com/intelligencer/article/openai-chatgpt-ai-cheating-education-college-students-school.html},
	language = {en},
	urldate = {2025-07-03},
	journal = {Intelligencer},
	author = {Walsh, James D.},
	month = may,
	year = {2025},
}

\end{document}